\documentclass[aps,prl,twocolumn,superscriptaddress,groupedaddress]{revtex4-2}

\usepackage{amsthm,amsmath}
\usepackage[utf8]{inputenc} 

\usepackage{graphicx}
\usepackage{dcolumn}
\usepackage{bm}
\usepackage{mathtools}
\usepackage[colorlinks=true]{hyperref}
\usepackage[dvipsnames]{xcolor}
\usepackage[mathlines]{lineno}

\newcommand\myshade{85}
\colorlet{mylinkcolor}{Red}
\colorlet{mycitecolor}{Blue}
\colorlet{myurlcolor}{Gray}

\hypersetup{
  linkcolor  = mylinkcolor!\myshade!black,
  citecolor  = mycitecolor!\myshade!black,
  urlcolor   = myurlcolor!\myshade!black,
  colorlinks = true,
}

\begin{document}

\title{The Fermionic Axion Interferometer}

\author{Nicolò Crescini}
	\email{nicolo.crescini@unipd.it}
    \affiliation{Dipartimento di Fisica e Astronomia "Galileo Galilei", Università di Padova, Italy}

\begin{abstract} 
The axion is an hypothetical beyond the Standard Model particle. Its experimental search is an ongoing effort, and an expanding number of techniques keep on narrowing its parameters space.
Leveraging the interaction between dark matter axions and spins, a fermionic interferometer is an experiment which aims at detecting the axion-induced precession of a spin resonance. 
We describe the detection scheme, outline the possible experimental implementations, their sensitive axion-mass range and discovery potential.
Furthermore, the building and characterization of an axion interferometer is explained in details and the resulting setup is used to search for sub-neV axion-like dark matter, obtaining a limit on its coupling to electrons.
\end{abstract}

\maketitle

\section{Axions and Spins}
\label{sec:intro}

The main evidences of physics beyond the Standard Model \cite{ELLIS2009187c} are the strong CP problem, the asymmetry between matter and anti-matter, neutrino oscillations, dark matter and dark energy \cite{doi:10.1146/annurev.nucl.012809.104433}.
The axion \cite{10.21468/SciPostPhysLectNotes.45} is an hypothetical particle which has the potential to solve two of these matters.
Originally proposed to accommodate for the absence of charge-parity symmetry violation in quantum chromodynamics, i.\,e. the strong CP problem \cite{DILUZIO20201}, it later became a compelling dark matter candidate \cite{doi:10.1126/sciadv.abj3618,BERTONE2005279}.
This theoretical sparkle triggered the first experimental searches that excluded early axion models, and which were followed by new models and new experiments \cite{IRASTORZA201889,RevModPhys.93.015004,doi:10.1126/sciadv.abm9928}. To this day, no signal compatible with axions has been reported.

The first thing to consider when detecting a particle is its mass, and the axion one is unknown. A virtually infinite mass range can be constrained by means of models \cite{PhysRevLett.118.031801}, astrophysics \cite{Giannotti_2017} and cosmology \cite{MARSH20161}, suggesting a preferred window for the axion mass $m_a$ between micro- and milli-electronvolts.
Still, many theoretical efforts and experimental searches are directed towards lighter or heavier axions or axion-like particles \cite{https://doi.org/10.48550/arxiv.2203.14923}.
Typically, the most sensitive experiments cover a narrow mass range, as in the case of haloscopes \cite{PhysRevLett.127.261803,Adair2022,PhysRevLett.130.071002,PhysRevD.106.052007,PhysRevLett.118.061302,MCALLISTER201767,PhysRevD.107.055013,grenet2021grenoble}, while broadband techniques cover a wide mass range at the expense of a reduced sensitivity \cite{PhysRevLett.127.081801,Gramolin2021,Anastassopoulos2017}. 
Among all of these are spin resonance axion searches can be based on electrons \cite{Crescini2018,FLOWER2019100306,PhysRevLett.124.171801}, or nuclei, which can access low frequency axions \cite{PhysRevLett.122.191302,doi:10.1126/sciadv.aax4539,10.1007/978-3-030-43761-9_13,Blanchard2023,Afach2021,Bloch2023}.

Dine-Fischler-Srednicki-Zhitnitsky (DFSZ) axions \cite{DINE1981199,Zhitnitsky:1980he} interact with fermions with pseudoscalar interaction constant $g_\mathrm{p}$ \cite{BARBIERI1989357,SUN2020135881} though the Lagrangian
\begin{equation}
    {\cal L}_f = \frac{g_\mathrm{p}}{2m_f} (\bar{\psi} \gamma_\mu \gamma_5 \partial^\mu a \psi),
    \label{eq:l_int}
\end{equation}
where $\partial^\mu a$ is the axion field quadriderivative, $\gamma_\mu$ and $\gamma_5$ are Dirac matrices, and $\psi$ is the quantum field of a fermion with mass $m_f$ and charge $e$.
One can notice that ${\cal L}_f$ is analogous to the interaction Lagrangian of quantum electrodynamics
\begin{equation}
    {\cal L}_\mathrm{QED} \supset -e (\bar{\psi} \gamma_\mu A^\mu \psi),
    \label{eq:l_int_qed}
\end{equation}
minus a $\gamma_5$. Physically, this means that the axion field derivative acts on fermionic spins as an effective magnetic field.
In the non-relativistic limit Eq.\,(\ref{eq:l_int}) can be recast in terms of the fermion's spin $s_i = \mu_f \sigma_i$, where $\mu_f = e/2m_f$ is the fermion magneton and $\sigma_i$ is the Pauli vector. The remaining terms form a pseudo-magnetic axionic field
\begin{equation}
    b_i = \frac{g_\mathrm{p}}{2e}\partial_i a,
    \label{eq:b_eff}
\end{equation}
which oscillates at the frequency of the axion mass $m_a$, such that $b_i \propto \beta_i e^{-i m_a t}$, where $\beta_i=(\beta,0,0)$ is the axion speed relative to the speed of light $c$.
If the spin $s_i$ is in a static magnetic field $B_i$ and in an axion effective field, both parallel to its direction, its equations of motion are Bloch equations
\cite{ashcroft2011solid}
\begin{equation}
    \dot{s_i}/\gamma = \epsilon_{ijk}s^j (B^k + b^k),
    \label{eq:b_eff_dynamics}
\end{equation}
where $\gamma$ is the fermion gyromagnetic ratio.
This equation displays a Larmor precession with an additional axion-induced spin precession \cite{PhysRevD.30.130,PhysRevD.92.105025,PhysRevD.103.116021}, which is the core observable of the present work.
Taking the non-relativistic limit of Eq.\,(\ref{eq:b_eff}), and treating the axion as a classical field are justified by $\beta_i \ll 1$ and by the large axion occupation number $n_a = \varrho_\mathrm{dm}/m_a$ \cite{BARBIERI2017135,PhysRevLett.124.171801}, respectively. An Earth-based laboratory is immersed in the dark matter halo of the Milky Way, whose estimated density is $\varrho_\mathrm{dm}=0.4\,\mathrm{GeV/cm^3}$, and travels through it at a speed $\beta\simeq10^{-3}$. The resulting axion effective field is
\begin{equation}
    b_a = \frac{g_\mathrm{p}}{2e}\beta n_a^{1/2} \simeq 3\times10^{-18}\left( m_a / 1\,\mathrm{eV} \right)\,\mathrm{T}.
    \label{eq:b_eff_numbers}
\end{equation}
centered at the frequency corresponding to the axion mass, and with a quality factor which is approximately $Q_a \simeq \beta^{-2} \simeq 10^6$ \cite{RevModPhys.93.015004,BARBIERI2017135}.

\begin{figure}
    \centering
    \includegraphics[width=.4\textwidth]{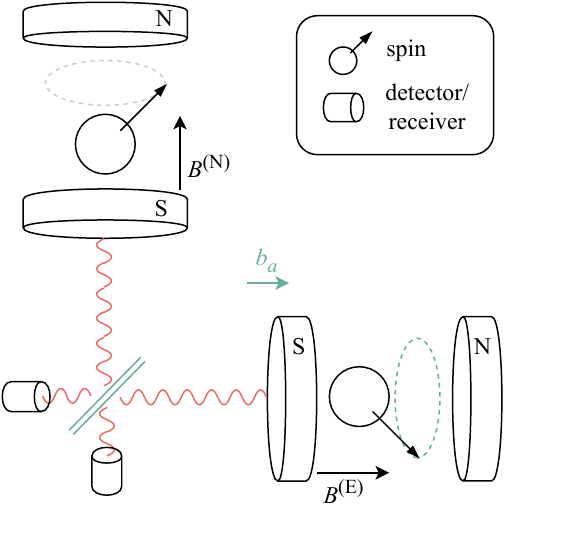}
    \caption{Scheme illustrating the principle of a fermionic interferometer. Two perpendicular spin resonances act as arms of an interferometer, which are probed by split light beams. The probe signals are reported in pink while the axion field $b_a$ is green, see text for further details.}
    \label{fig:general_scheme}
\end{figure}

The detection of an axion precession can be understood in a general experimental scheme which can be realized at different energy scales---i.e. axion masses---and therefore with different physical systems and technologies, outlined in Table \ref{tab:fermionic_interferometers}.
Let's first discuss the experimental scheme, outlined in Fig.\,\ref{fig:general_scheme}. The spin resonance under consideration is phase-modulated by the dark matter axion field gradient which, being dependent on $\beta_i$, is a directional effect. This suggests that two spin resonances with static magnetic fields oriented perpendicularly to each other can be used as the two arms of an interferometer. The east arm---(E)---is parallel to the axion field gradient and carries the dark matter signal, while the north arm---(N)---acts as a reference.
In the east arm $B^\mathrm{(E)}_i = (B_0,0,0)$ and $b_i = (b_a,0,0)$, so Eq.\,(\ref{eq:b_eff_dynamics}) describes a spin resonance whose resonant frequency $\omega_0 = \gamma B_0$ is periodically modulated by an axionic field with amplitude $b_a$ and frequency $\omega_a$. The north arm has $B^\mathrm{(N)}_i = (0, B_0, 0)$, giving the same resonance frequency and no axionic modulation.
One can therefore imagine to shine the same light beam to both the resonances, and then let the outputs interfere with themselves, providing an interferometric readout of a dark matter axion signal.

\begin{table}[b]
\begin{tabular}{ c | c c c }
                        & $\omega_0/2\pi$     &   $k_0/2\pi$   &   $m_a$ \\ 
\hline
Spin resonance          & 10\,MHz   &   100\,kHz       &   $<0.4$\,neV         \\  
Magnon polariton        & 10\,GHz   &   1\,MHz         &   $<40$\,neV          \\ 
Atomic transition     & 100\,THz  &   10\,GHz        &   $<40$\,$\mu$eV      \\
\end{tabular}
\caption{Experimental platforms to implement a fermionic axion interferometer at different energy scales.}
\label{tab:fermionic_interferometers}
\end{table}

This scheme is analogous to the one of gravitational wave observatories, where the Fabry-Pérot cavities are substituted by the spin resonances, as shown in Fig.\,\ref{fig:general_scheme}.
When a photon interacts with a resonant system, a cavity or a spin-resonance, the optical path of the photon is increased by its lifetime in the resonance, related to the number of reflections on the cavity walls or to the coherence time of the spin, in turn increasing the effect of gravitational waves or of the axion field.
Although existing proposals use interferometric techniques to search for axions \cite{PhysRevD.98.035021,PhysRevLett.121.161301,PhysRevD.100.023548,PhysRevD.101.095034}, the present one differs from them in both detection principle and practical realization.
As is detailed in the following, the advantages of this approach are that it leverages the considerable experimental assets of interferometry, it is broadband in the axion mass parameter space, and it can be engineered at different energy scales.

An interferometric readout suppresses its arms common noise. Possible sources are mainly environmental electromagnetic disturbances, which affect all the experimental configurations of Table\,\ref{tab:fermionic_interferometers}, while specific systematic errors depend on the implementation, for instance, cross-talks at microwave frequencies or vibrations at optical frequencies.
The fundamental limit of this detection scheme is expected to be quantum shot noise of the photon beam \cite{PhysRevLett.125.131101}, which poses a cap on the signal-to-noise increase obtainable intensifying the probe.

\section{Detection Scheme}
\begin{figure*}
    \centering
    \includegraphics[width=.93\textwidth]{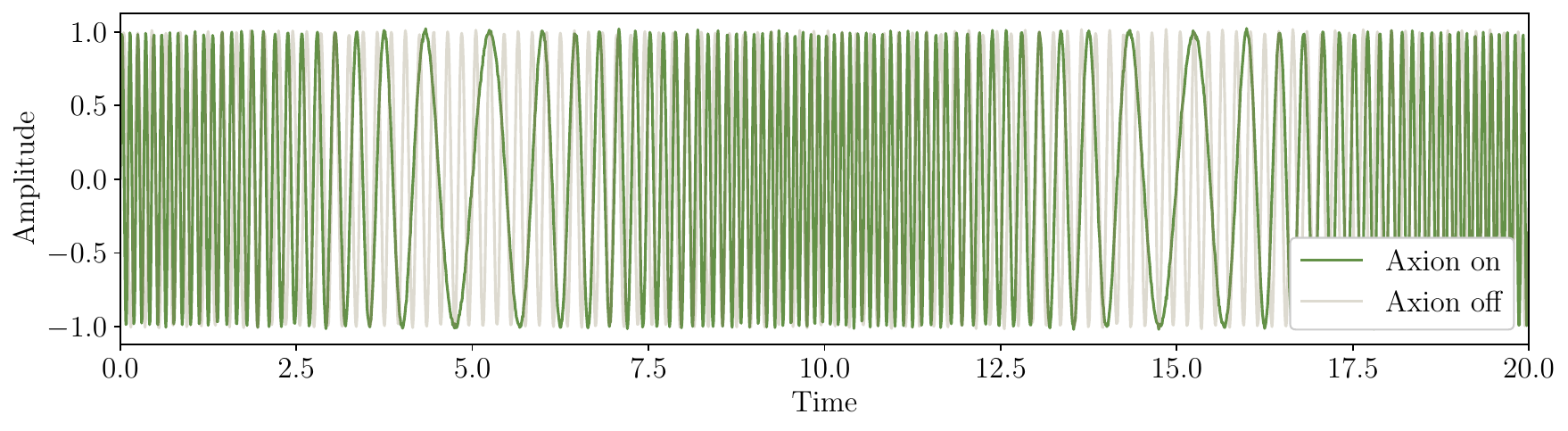}
    \caption{Effect of the axionic modulation on a monochromatic signal interacting with a spin resonance. The axionic dark matter modulates the spin resonance frequency, thereby inducing oscillations on the phase of the monochromatic signal.}
    \label{fig:axion_effect}
\end{figure*}

As shown in the introduction, axionic dark matter produces coherent oscillations in the frequency of a spin resonance. A suitable way to detect this effect is based on phase modulation, and is detailed in the following.
A monochromatic input light on-resonance with the spin precession and with amplitude $A_0$ is phase modulated by the oscillating resonance, yielding a signal
\begin{equation}
    \xi(t) = A_0 e^{-i\omega_0 t} e^{-ix\sin(\omega_a t)},
    \label{eq:phase_modulation}
\end{equation}
where $x=2\pi^2 \gamma b_a / k_0$ is the modulation index, and $k_0$ is the spin resonance linewidth. 
This effect is illustrated in Fig.\,\ref{fig:axion_effect} by using a magnified effect of axions acting on a frequency $\omega_0/2\pi = 5\,\mathrm{MHz}$.
If the modulation index $x$ is small, $\xi(t)$ can be recast in terms of the first kind Bessel functions
\begin{equation}
    \xi(t)  = \sum_{n=0}^{\infty}\xi_n(x) 
            = A_0 e^{-i\omega_0t} \sum_{n=0}^{\infty} \frac{x^n}{2^n n!}e^{in\omega_at}.
    \label{eq:bessel}
\end{equation}
At zero order Eq.\,(\ref{eq:bessel}) gives the monochromatic tone amplitude itself $|\xi_0| = A_0$, while for $n=1$ it yields the first sideband amplitude
\begin{equation}
    |\xi_1| = \pi \gamma A_0 b_a / 2k_0
    \label{eq:sideband}
\end{equation}
which is the searched-for signal of the scheme. One can notice that the sensitivity of phase-modulation schemes to first order does not depend on the amount of material used, but rather on its coherence properties---i.e. the spin resonance linewidth $k_0$---making them ideal table-top experiments \cite{PhysRevApplied.16.034036}.

Eq.\,(\ref{eq:sideband}) needs to be compared with a background to get the experimental sensitivity. While the signal calculation is general enough to accommodate for different experimental schemes, the treatment of the noise is more involved, and is discussed hereafter in some details.
Assuming that the sensitivity is limited by a white noise with power spectral density $N^2_0$, we can calculate the minimum detectable effective axion field as
\begin{align}
\begin{split}
    \sigma_a &= \frac{2 k_0 N_0}{\pi \gamma A_0} \\
             &= 2.27 \left(\frac{k_0}{10\,\mathrm{kHz}}\right) \left(\frac{N_0/A_0}{10^{-5}/\mathrm{\sqrt{Hz}}}\right)\,\frac{\mathrm{pT}}{\mathrm{Hz}^{1/2}},
    \label{eq:b_min}
\end{split}
\end{align}
where $\gamma$ is the electron gyromagnetic ratio, and the other quantities are roughly the experimental parameters of the setup described in the next part of this work.
For instance, considering an axion signal at 1\,kHz and therefore an integration time of $10^3\,\mathrm{s}$---matching $Q_a$---we obtain a sensitivity of about 72\,fT. 

The interferometer bandwidth---the range of axion masses probed by the experiment---is in first approximation $k_0$, as when $\omega_a>k_0$ the sideband $\xi_1$ falls out of resonance, and its amplitude decreases. However, depending on the background origin, the signal-to-noise ratio could be preserved, and the experimental bandwidth could be increased depending on the measurement technique.

Let's briefly discuss the three experimental cases of Table \ref{tab:fermionic_interferometers}, which are referred to as radiofrequency, microwave, and laser interferometer, respectively.
A radiofrequency interferometer can be realized with both nuclei or electrons. The former resembles some configurations of the cosmic axion spin precession experiment \cite{PhysRevD.88.035023,PhysRevX.4.021030}, although not searching for axion-induced oscillating dipole moments or magnetization. However, one could use a different detection configuration of the same setup to implement the scheme described in this work.
Electron spin based experiments work with low magnetic fields, which is advantageous in terms of complexity but makes it particularly sensitive to environmental fluctuations. 
A non-negligible advantage of radiofrequency setups is that all the interferometry can be handled digitally thanks to fast electronics, reducing the hardware requirements to a minimum.
At microwave frequencies nuclear spin resonance becomes inaccessible, although some nuclei like Holmium show up to GHz precession \cite{DStPBunbury_1985,DStPBunbury_1989} thanks to a large effective $\gamma$.
Electron spin resonances on the other hand are widely studied in this frequency range. A major problem occurring with spin resonances at high frequencies is the broadening of the resonance linewidth, often referred to as radiation damping \cite{BARBIERI2017135}, which is eliminated by embedding the sample in a resonant cavity. The resulting photon-magnon hybrid systems are applied e.\,g. to quantum technologies \cite{PhysRevLett.113.083603} and Dark Matter searches \cite{PhysRevLett.124.171801}.

\begin{figure}[b]
    \centering
    \includegraphics[width=.4\textwidth]{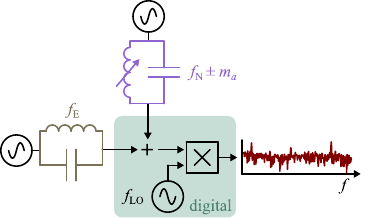}
    \caption{Circuit scheme of the halocope. The two resonators represent the ferrimagnetic arms of the setup, which are probed with a coherent tone each. The resulting signal is acquired with a data acquisition board and processed.}
    \label{fig:circuit_scheme}
\end{figure}

\begin{figure*}
    \centering
    \includegraphics[width=.96\textwidth]{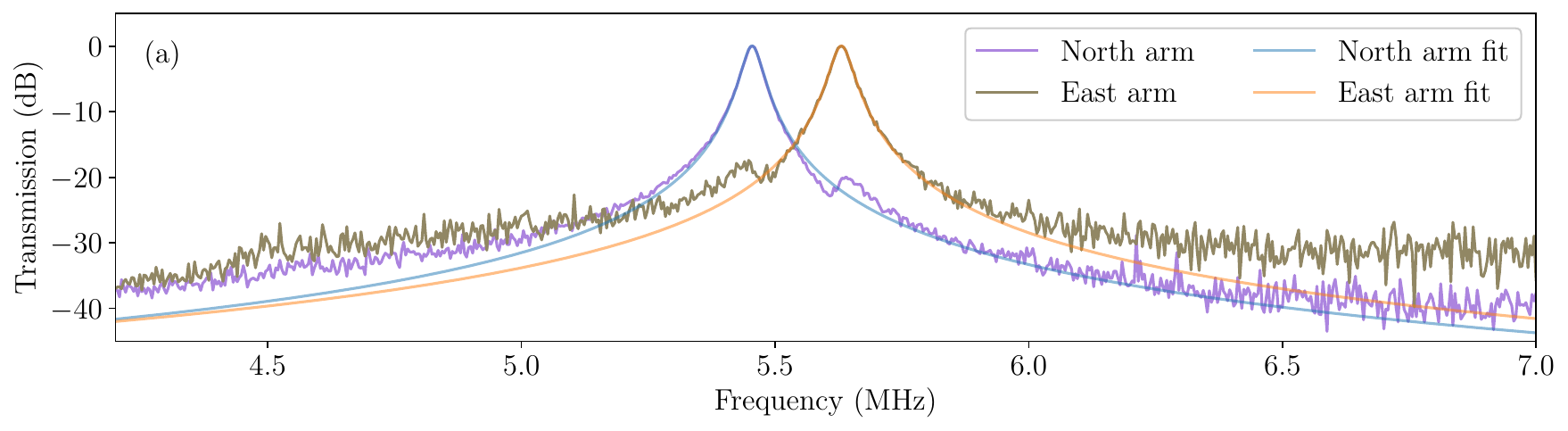}
    \includegraphics[width=.96\textwidth]{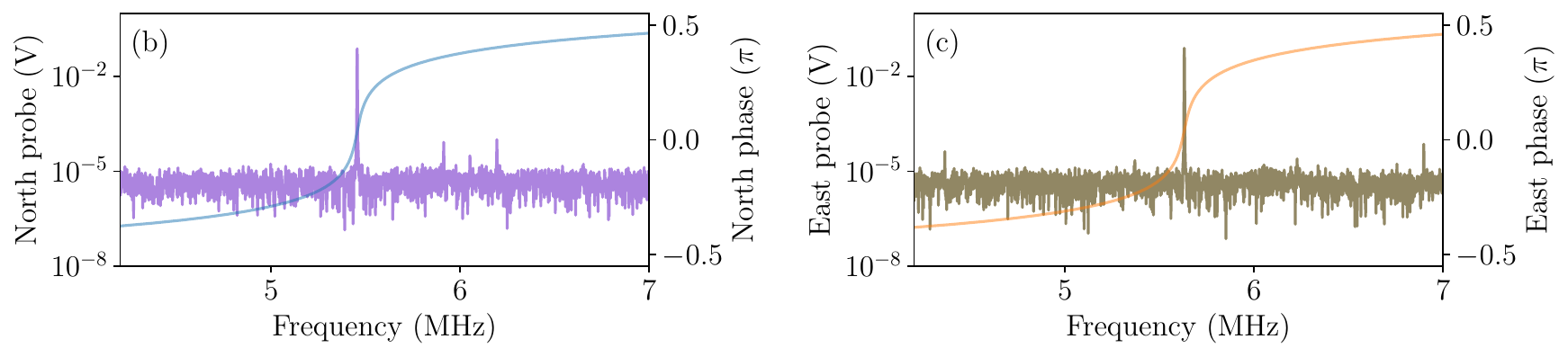}
    \caption{Characterization and operation of the fermionic interferometer. Panel (a) shows the transmission spectrum of the north and east resonances used to calculate the experiment properties and set the tones frequencies $\omega_1$ and $\omega_2$. Panels (b) and (c) are an extract of the interferometer data while in operation, and show the probing tones superimposed to resonance phases.}
    \label{fig:operation}
\end{figure*}

A laser axion interferometer has several intriguing features. 
Let us consider as an example a magnetic dipole atomic transition in the infrared range. A resonant laser beam transmitted through the sample will experience phase modulation like the carrier signals treated beforehand, and with a quality factor of order $10^4$ the bandwith of the interferometer would encompass the whole QCD axion range \cite{AxionLimits}.
On top of that, the experimental apparatus can benefit from all the advantages of interferometers, similarly to gravitational waves observatories. However,  the effective modulation of a magnetic dipole resonance should be estimated and would probably depend on the nature of the transition itself \footnote{The effective amplitude modulation and the noise estimation are not trivial, therefore a thorough discussion of this setup will be carried out in a forthcoming work}.

\section{Experimental Axion Search}
\label{sec:kakapo}

This section concerns the building and operation of a radiofrequency axion interferometer sensitive to the axion-electron interaction at sub-megahertz frequency \cite{openhaloscope}, a scheme of the experimental setup is presented in Fig.\,\ref{fig:circuit_scheme}. Given the corresponding mass range, this measurement is configured as an axion-like particle search.

The sensing material is a ferrimagnet, NiZn, shaped in two rods oriented perpendicularly.
The rods are the two arms of the interfetometer and are labeled north (N) and east (E) from the direction they are pointing at. Each rod is readout by coupling it to a coil, resulting in a resonance frequency of about 5\,MHz, and is then connected to the electronics (for more details see the appendix).

\begin{figure*}
    \centering
    \includegraphics[width=.96\textwidth]{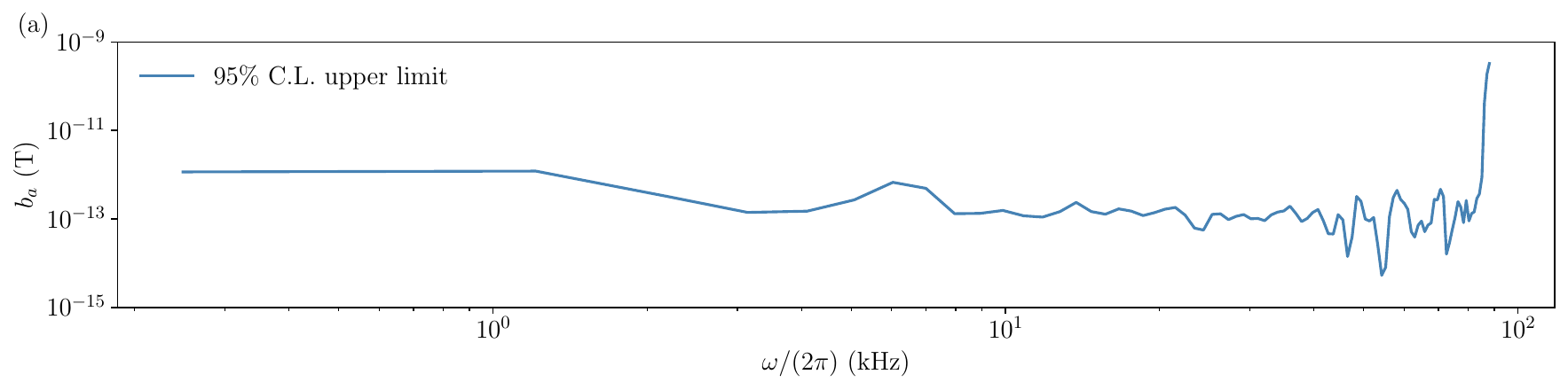}\\
    \includegraphics[width=.96\textwidth]{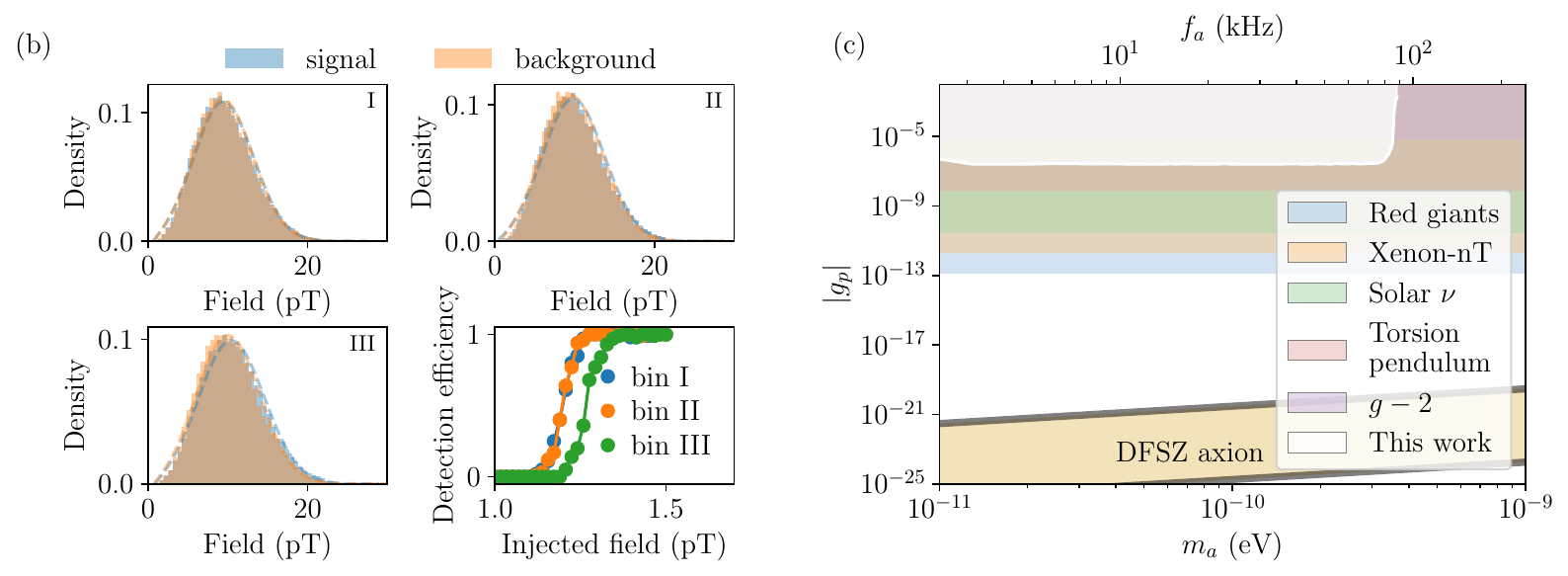}
    \caption{Run results. (a) 95\% confidence-level upper limit on the axion-induced effective magnetic field $b_a$ obtained from the interferometric search. (b) Histograms of the measured field amplitude for representative frequency bins, showing the signal and background distributions, together with the corresponding detection efficiency as a function of injected field amplitude, used to calibrate the sensitivity of the analysis in each bin. Bins I and II are representative noise bins, while bin III corresponds to a candidate excess in the field amplitude; this candidate did not pass the subsequent vetting procedure and was therefore not an axion signal. (c) Resulting exclusion limit on the axion-electron coupling $|g_p|$ as a function of axion mass $m_a$. Exclusion plots and axion line \cite{AxionLimits}. Existing constraints are shown from astrophysical red-giant cooling \cite{PhysRevD.102.083007} and solar neutrino searches \cite{PhysRevD.79.107301,Gavrilyuk2022}. Laboratory searches include Xenon-nT \cite{PhysRevLett.129.161805}, torsion-pendulum experiments \cite{PhysRevLett.115.201801}, and muon $g-2$ measurements \cite{Yan2019}. The shaded band indicates the DFSZ axion model prediction, and the white region shows the parameter space excluded in this work at 95\% C.L.}
    \label{fig:exclusion_plot}
\end{figure*}

Given the relatively low frequency of the resonances, the signal is generated and readout by a 125\,MS/s analog-to-digital and digital-to-analog field-programmable-gate-array board, allowing to handle the signal processing digitally. In addition, several environmental parameters, like temperature, pressure or local magnetic field, are recorded with dedicated sensors, allowing to disentangle a possible axion signal from environmental systematics.
The setup characterization relies on the transmission measurement of the resonances, which gives the resonance frequencies of the arms, and their linewidth, as presented in Fig.\,\ref{fig:operation}a. The north arm has resonance frequency $\omega_\mathrm{N} = 5.45\,\mathrm{MHz}$ and linewidth $k_\mathrm{N} = 19.4\,\mathrm{kHz}$, while the east arm has $\omega_\mathrm{E} = 5.63\,\mathrm{MHz}$ and $k_\mathrm{E} = 21.4\,\mathrm{kHz}$, obtained by fitting the transmission measurement with Lorentzian functions.

The haloscope operation requires to send two tones on resonance with each arm to achieve the maximum phase modulation, as shown in Fig.\,\ref{fig:operation}b and \ref{fig:operation}c. 
The tones' frequency are $\omega_\mathrm{N}$ and $\omega_\mathrm{E}$, while their amplitudes are set to $A_\mathrm{N}=A_\mathrm{E}\simeq10.0\,\mathrm{V}$. The two analog signals are summed and down-converted according to Fig.\,\ref{fig:circuit_scheme}, where the down-conversion frequency $\omega_\mathrm{LO}=(2\pi)f_\mathrm{LO}$ needs to be $\omega_\mathrm{LO}=(\omega_\mathrm{N} + \omega_\mathrm{E})/2$. The interferometer can suppress common phase or amplitude noise depending on the phase of the down-conversion tone \cite{PhysRevApplied.20.044072}, and given the nature of the axion signal, we opt for phase noise reduction.
Eq.\,(\ref{eq:b_min}) shows that it is possible to arbitrarily improve the sensitivity by increasing $A_0$, i.\,e. using stronger tones. However, in a realistic experimental configuration $N_0$ is directly proportional to $A_0$, or---in the best case scenario---dominated by quantum fluctuations.
The limit of the present setup is the phase noise of the signal generators, which is about -100\,dBc, and fixes $A_0/N_0\simeq10^{-5}\,\mathrm{V/Hz^{1/2}}$.
As mentioned in the introduction, the virialized axion field have $Q_a\simeq10^6$, and therefore the optimal resolution bandwidth to detect it is $\mathrm{RBW}=\omega_a/Q_a$. For frequencies below 200\,kHz $\mathrm{RBW}$ is above one second, but in the present setup the resolution bandwidth is set to $\mathrm{RBW}=1\,\mathrm{kHz}$, the minimum allowed by the electronics with a decimation factor of 8.
The duty cycle of the acquisition is also limited to roughly 2\%. These last two limitations are purely technical, and can be optimized in an updated version of the setup.
The experiment is controlled using a dedicated characterization, acquisition and analysis code which can be found in Ref.\,\cite{openhaloscope}.

If one arm is oriented towards the dark matter wind \cite{BARBIERI2017135} its resonance frequency is fully modulated by the axion field $b_a$, while the one of the remaining arm is unaltered.
A single radiofrequency signal probes the on-resonance transmission of the ferrite, and indicates whether the resonance frequency shifts, while two signals, each on one arm of the interferometer, can detect relative shifts of one spin resonance with respect to the other \cite{PhysRevApplied.20.044072}. 
The output data (see Fig.\,\ref{fig:circuit_scheme}) is a spectrum which contains information on whether $\omega_\mathrm{N}$ shifted with respect to $\omega_\mathrm{E}$ or viceversa, which can be used to put a limit on the presence of a dark matter wind.
The pilot experimental run consists in two acquisitions which are referred to as signal and background. The setup is mounted on an astronomical computerized mount aiming towards Vega, which keeps the interferometer north arm aligned to the Dark Matter wind \cite{BARBIERI2017135} for the signal run. The background run is collected with the same setup rotated by 90 degrees, such that both arms are unaffected by the wind \footnote{Depending on the detection and analysis strategy, this configuration can be improved and was chosen for the pilot run mostly for its simplicity.}.

The result of the run are shown in Fig.\,\ref{fig:exclusion_plot}.
Each acquisition consists of $10^3$ data blocks, each containing $30$ statistically independent $1\,\mathrm{ms}$-waveforms, totaling $N=30\,000$. The power spectrum is estimated via a Blackman-windowed periodogram, converted to magnetic field, and saved into a single spectral matrix $S_{ij}$ ($i$ frequency bins, $j$ waveform). An identical matrix $B_{ij}$ is produced from the background run. 
The per-bin field excess $\Delta_i = \langle S_i \rangle - \langle B_i \rangle$  and its uncertainty $\sigma_i = \sqrt{[\mathrm{Var}(S_i) + \mathrm{Var}(B_i)]/N}$ are estimated from the sample variances over the waveform ensemble. 
Although the individual field amplitudes follow Rice statistics (Fig.\,\ref{fig:exclusion_plot}b), averaging over $N=30\,000$ independent waveforms yields an excess that is well approximated by a Gaussian distribution according to the central limit theorem.
Bins with $\Delta_i > 3\sigma_i$ are vetted by requiring the excess to be positive, not driven by outlier blocks, and isolated to a single bin, while surviving candidates are referred to a follow-up measurement.

The complete analysis pipeline is validated through bootstrap signal-injection studies \cite{PhysRevLett.127.261803}, in which a synthetic narrowband signal of known strength is added to background realizations and the recovery fraction is used to determine the detection efficiency.
No candidate passed all criteria, consistent with the absence of an axion-like signal. 
A $95\%$ Bayesian upper limit $\Lambda_i$ is obtained from the upper edge of the
credible interval of a Gaussian posterior truncated at $s_i \geq 0$,
\begin{equation}
p(s_i \mid \Delta_i,\sigma_i)=
\begin{cases}
\dfrac{\exp\!\left[-\dfrac{(s_i-\Delta_i)^2}{2\sigma_i^2}\right]}
{\mathcal N_i\sqrt{2\pi}\,\sigma_i},
& s_i\ge0,\\[2mm]
0, & s_i<0.
\end{cases}
\label{eq:truncated_posterior}
\end{equation}
where $\mathcal{N}_i$ normalizes the posterior over $s_i \ge 0$.
The upper limit $\Lambda_i$ is then defined by the credible-interval
condition
\begin{equation}
    \int_0^{\Lambda_i} p(s_i \mid \Delta_i, \sigma_i)\, ds_i = \mathrm{C.L.},
    \qquad \mathrm{C.L.} = 0.95.
    \label{eq:upper_limit_def}
\end{equation}
The resulting upper limit is converted into a constraint on  using
Eqs.\,(\ref{eq:b_eff_numbers}) and (\ref{eq:sideband}) \cite{PhysRevD.101.123011}, and is reported in terms of effective field $b_a$ in Fig.\,\ref{fig:exclusion_plot}a and of coupling constant $g_\mathrm{p}$ in Fig.\,\ref{fig:exclusion_plot}c.

\section{Perspectives and Conclusions}
\label{sec:conc}
A fermionic interferometer is a broadband dark matter axion experiment based on the modulation of spin resonances. Two light beams interact with a magnetic material whose properties depend on the local dark matter density to then interfere, eliminating systematic errors and only preserving the searched-for signal. The scheme can be realized with radiofrequency, microwave or laser light.
We present the experimental implementation and operation of a radiofrequency fermionic interferometer with which we perform an axion-like particle search in the sub-neV mass range by testing its coupling to electrons.

This pilot setup of the interferometer is a minimal experimental scheme which is prone to improvements, and which is simple enough to be openly reproduced \cite{openhaloscope}. Short term improvements include an apparatus where the interferometry is analog, and digitization happens only after the down-conversion. This allows for continuous monitoring of the signal, and therefore an arbitrary resolution bandwidth and a unitary duty cycle, making it also possible to search for transient signals.

For future axion searches, in particular towards $\mu\mathrm{eV}$-mass axions (see Table \ref{tab:fermionic_interferometers}), one can envision a laser interferometer probing an infrared magnetic resonance of e.\,g. Er:YLF or Er:YSO \cite{MACFARLANE20021,Kukharchyk_2018,PhysRevB.79.115104}. This axion experiment is particularly interesting because it leverages the analogy with gravitational waves detectors, where the magnetic resonance plays the role of the Fabry-Pérot cavity. However, this experiment design and operation requires a detailed study of the atomic transitions to be used.

Beyond axion searches, an intriguing possibility is the use of this interferometer to detect gravitational waves effects via gravitomagnetic effects \cite{ctx9998987570006046} or gravitational precession \cite{PhysRevD.95.104044,PhysRevD.102.101501}. No change is necessary on the experimental setup, as it is already sensitive to any effective magnetic field modulating the spin resonance, but the analysis procedure would need to be modified in order to measure transient effects instead of persistent ones.
Eventually, we mention that the magnetic field sensitivity of the setup is interesting also beyond fundamental physics, as when optimized can compete with state-of-the-art magnetometers \cite{PhysRevApplied.16.034036,PhysRevApplied.19.044044,PhysRevLett.130.063201,clarke2006squid,PhysRevApplied.18.L021001}.

\section{Data availability}
\label{sec:data}
The data used in this work are publicly available on Zenodo \cite{openhaloscope-zenodo} while all the code to operate the interferometer and analyze the data can be found in the repository of Ref.\,\cite{openhaloscope}.

\section{Acknowledgements}
\label{sec:ack}
The support of Amberlab, in the persons of Davide Fasoli and Manuel Pachera, is greatly acknowledged for the help building the experimental apparatus. 
Federico Chiossi is also acknowledged for the advice on the use of atomic transitions to realise the fermionic interferometer. Eventually, the author would like to thank Gianni Carugno and Giuseppe Ruoso for the discussion on the experimental scheme.

\bibliographystyle{apsrev4-2}
\bibliography{ferm_ax}

\appendix
\section{Appendix A: Signal Extraction}
\label{app:signal}
This appendix briefly discusses a possible analysis routine to be implemented in a more advanced version of the experiment. The non-continuous acquisition of this prototype is a strong limitation which, however, can can be overcome with an improved and still inexpensive setup composed of two oscillators, a mixer, and a low sampling-rate digitizer, as mentioned in the conclusions.

The present analysis routine is minimal, aiming only to establish an upper limit, but continuous signal acquisition would enable a more refined time-domain analysis. Let's consider a time trace of length $T$, in this scenario the resolution bandwidth can be optimized for the test frequency $\omega_i$, for instance, by splitting acquired data into $N_i = T/t_i$ arrays of length $t_i = Q_a/\omega_i$ and fitting them with a sine wave of frequency $\omega_i$. 
The resulting sine wave amplitude $A_i$ corresponds to the Fourier component of the data at that frequency \cite{press_etal:1992}, and can be averaged $N_i$ times. This process is then repeated at $\omega_{i+1} = \omega_i + 1/t_i$, yielding $A_{i+1}$. Eventually, these amplitudes form a spectrogram in which transient signals or, particularly relevant for axion-like dark matter, daily modulations can be identified. The measured signal can be fitted to the daily modulation to derive an upper limit at the specified frequency, or alternatively, given the compactness of the apparatus, one can modulate its orientation and thereby the signal at a desired frequency.

\section{Appendix B: Open Source Project}
\label{app:open}

\begin{figure}[b]
    \centering
    \includegraphics[width=.35\textwidth]{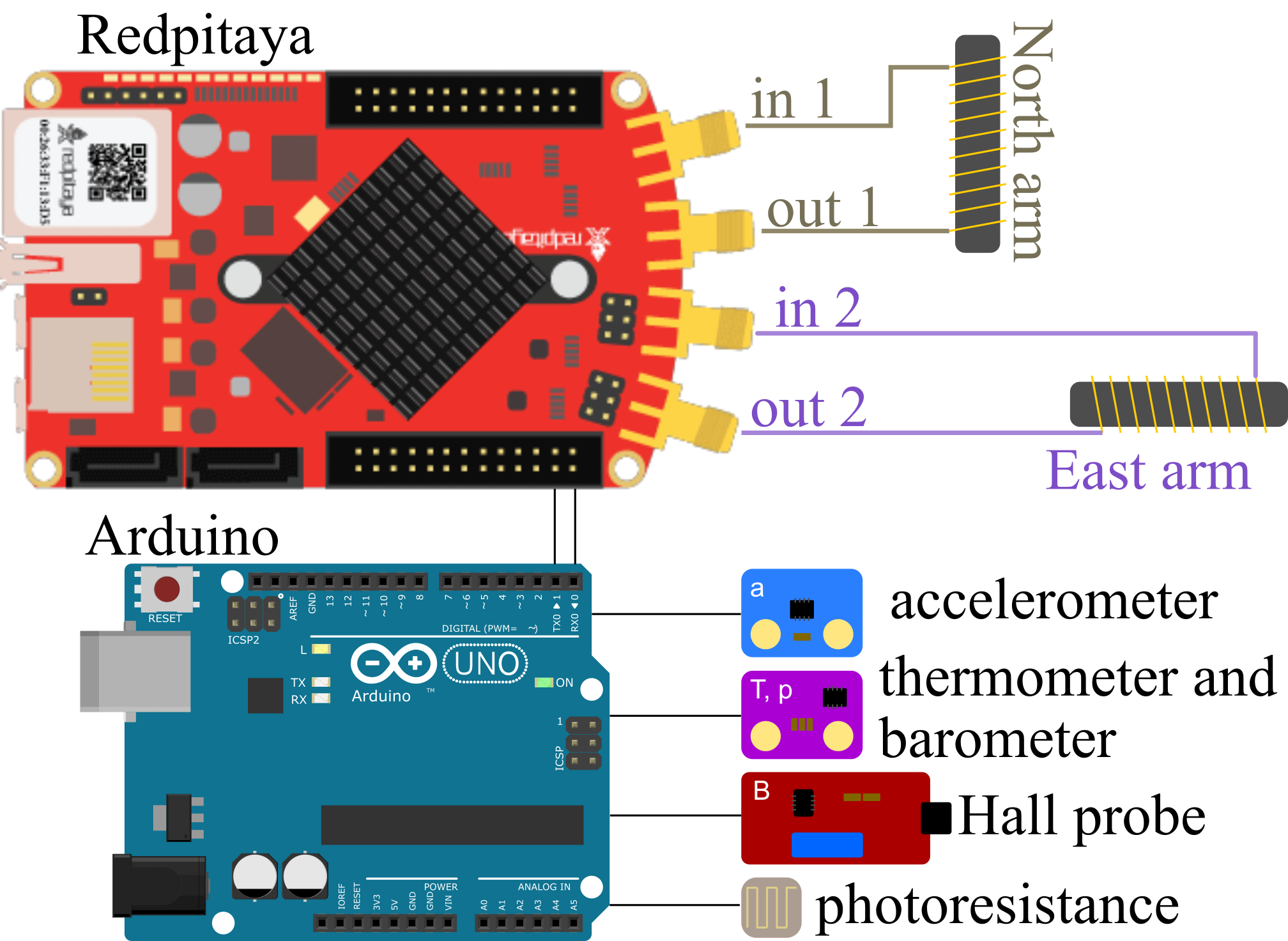}
    \caption{Circuit board scheme of the haloscope.}
    \label{fig:setup_board}
\end{figure}
The simple hardware requirements of the experiment make it interesting for open source projects to be carried out in schools, universities or by amateurs. The described experimental setup is inexpensive and all of the necessary code, from drivers to analysis routines, is publicly available \cite{openhaloscope}.
A circuit-board scheme of the experiment is sketched in Fig.\,\ref{fig:setup_board}.

\end{document}